\documentclass[11pt]{article}
\usepackage{latexsym}
\usepackage{epsfig}
\usepackage{amsmath}
\usepackage{graphics}
\usepackage{amssymb}
\usepackage{amsthm}
\usepackage{amsfonts}
\usepackage{amsopn}
\usepackage{amscd}
\usepackage{fullpage}

\newtheorem{thm}{Theorem}

\newtheorem{defn}[thm]{Definition}

\numberwithin{equation}{section} 
\numberwithin{figure}{section}
\numberwithin{thm}{section}
\begin{document}
\pagenumbering{arabic}
\begin{center}
\textbf{A Comparison of Secret Sharing Schemes Based on Latin Squares and RSA}\\
\textbf{Liam Wagner\footnote[1]{Corresponding address: \\Department of
Mathematics The University of Queensland, St Lucia 4072 Qld Australia 
 Email: LDW@maths.uq.edu.au}\\Department of Mathematics and\\
St John's College within\\ The University of Queensland}
\end{center}

\begin{abstract}
\noindent In recent years there has been a great deal of work done on secret sharing
schemes. Secret Sharing Schemes allow for the division of keys so that an
authorised set of users may access information. In this paper we wish to 
present a critical comparison of two of these Schemes based on Latin 
Squares \cite{cpr} and RSA \cite{shoup}. These two protocols will be examined 
in terms of their positive and negative aspects of their security.
\end{abstract}


\section{Introduction}
In communications networks which require security, it is important that 
secrets be protected by more than one key. Furthermore a system of several 
keys with more than one way for their combination may allow for the unique 
recovery of a secret. Schemes that have a group of participants that could 
recover a secret are known as \textit{Secret Sharing Schemes}.
\\
The idea of secret sharing is to start with a secret, divide it into pieces
called \emph{shares}, which are then distributed amongst users such that the
pooled shares are specific subsets of users allowed to reconstruct the
original secret, \cite{men}.
\\\\
\textbf{Threshold Schemes}\\
\noindent Shamir \cite{sham}, describes threshold schemes as being very helpful in the
management of cryptographic keys. The most secure key management scheme keeps the key in a single place. This sort of scheme may not
always be appropriate, and an obvious solution to this may be to make multiple
copies of the key. This may increase the risk associated in keeping
multiple keys secret. By using Shamir's \cite{sham} threshold scheme concept we
can get a very robust key management scheme. 
\\
Threshold schemes are well suited to applications in which a group of
individuals with conflicting interests must cooperate \cite{sham}. By following
Shamir's \cite{sham} protocol and choosing the correct $t$ and $w$ parameters we can give any
sufficiently large majority the authority to take some action while giving any
sufficiently large minority veto powers. We shall now use the definition outlined in \cite{stin} to describe 
what a threshold secret sharing scheme is.
\begin{defn}
Let $t$ and $w$ be positive integers, $t\leq w$. A $(t,w)-$threshold scheme is
a method of sharing a key $K$ among a set of $w$ players (denoted by
$\mathcal{P}$), in such a way that any $t$ participants can compute the value of
$K$, but no group of $t-1$ participants can do so. 
\end{defn}
\noindent The value of $K$ is chosen by a special participant which is referred to by
\cite{stin} as the $dealer$. The dealer is denoted by $D$ and we must assume
that $D \notin \mathcal{P}$. When $D$ wants to share the key $K$ among the
participants in $\mathcal{P}$, $D$ gives each participant some partial information
referred to earlier as a share. The shares should be distributed secretly, so no
participant knows the share given to any other participant.
At some later time, a subset of participants $B \subseteq \mathcal{P}$ will pool
their shares or return them to the dealer in an attempt to compute the key $K$.
If $|B| \geq t$, then they should be able to compute the value of
$K$ as a function of the shares they collectively hold. Furthermore if $|B| <
t$, then they should not be able to compute $K$. If we follow the notation of
Stinson \cite{stin},
\begin{equation}
\mathcal{P} = \{P_{i}: 1 \leq i \leq w \}
\end{equation} as the set of participants, $\mathcal{K}$ is the set of keys and
$\mathcal{S}$ as the set of secrets. A useful point proposed by Shamir \cite{sham} is that a hierarchical scheme may
be created, so that some players may have shares which are of more importance 
(weight).

\subsection{Access Structures}
In our outline of threshold schemes, we wanted $t$ out of $w$ players to be able to
determine the key. A more general situation is to specifically exactly which subsets
of players should be able to determine the key and those that should not
\cite{stin}. If we describe $\Gamma$ as being a set of subsets of $\mathcal{P}$, and the
subsets in $\Gamma$ as being the subset of players that should be able to
compute the key. $\Gamma$ is denoted as being the access structure and the
subsets in $\Gamma$ are called authorised subsets. 
\\
Furthermore if we let $\mathcal{K}$ be the set of keys and $\mathcal{S}$ be the share set.
We shall continue to use the dealer $D$ who wants to share a key $k\in
\mathcal{K}$, and then gives each player a share $S\in \mathcal{S}$.
Some time later a subset of players will attempt to determine $K$ from the
shares they collectively hold. If we notice that a $(t,w)$-threshold scheme creates the access structure $\{B
\subseteq \mathcal{P}| \; |B| \geq t\}$, which is referred to by
Stinson \cite{stin} as the \emph{threshold access structure}.
\\
If $\Gamma$ is an access structure, then $B \in \Gamma$ is a minimal authorized
subset and $A \notin \Gamma$ whenever $A \subseteq B, A \neq B$. The set of
minimal authorized subsets of $\Gamma$ is denoted by $\Gamma_{0}$ and is called
the basis of $\Gamma$. Since $\Gamma$ consists of all subsets of $\mathcal{P}$
that are supersets of a subset in the basis $\Gamma_{0}$. Thus $\Gamma$ is
determined uniquely as a function of $\Gamma_{0}$ such that:
\begin{equation}
\Gamma = \{C \subseteq \mathcal{P}, \; B \subseteq C, \; B \in \Gamma_{0}\}
\end{equation}

\section{Latin Squares}
In their 1994 paper Cooper, Donovan and Seberry \cite{cpr} laid the foundation
for the use of critical sets as a combinatorial structure which could be used to
construct a secret sharing scheme. We should begin this section by defining a
Latin Square and the concept of a critical set.
\begin{defn}
A $n \times n$ Latin Square is an $n \times n$ matrix whose entries are taken
from a set of $n$ objects so that no object occurs twice in any row or column.
\end{defn} 

\begin{defn}
A critical set of a Latin Square L defined over the set $X =\{ 1,\dots,n\}$
where,
\begin{equation}
C = \{(i,j,k) \in X \times X \times X\}
\end{equation}
such that L is the only square of order n with $i$ in the $(j,k)$th for every
$(i,j,k) \in C$. Furthermore no proper subset of C may satisfies this condition 
\end{defn} 
\noindent An important construction which we need to define is the
concept of a strong critical set for a Latin Square. 
\begin{defn}
A critical set L is a strong critical set if there exists a set
$\{P_{1},\dots,P_{m}\}$ of $m=n^{2} - \|A\|$ partitions of order n, which satisfy
the following properties:
\begin{itemize}
\item $L \supset P_{m} \supset P_{m-1} \supset \dots \supset P_{2} \supset P_{1}
=A$
\item $\forall$ $i$, $1\leq i \leq m-1$, $P_{i} \cup \{(r_{i},c_{i},e_{i})\} =
P_{i+1}$ 
\item $P_{i} \cup \{(r_{i},c_{i},e_{i})\}$ is not a partial Latin Square such
that $\nexists e \in N \ \{e_{i}\}$
\end{itemize} 
\end{defn}

\begin{defn}
A critical set is referred to as being semi-strong, if there exists a set 
$\{P_{1},\dots,P_{m}\}$ of $m=n^{2} -|A|$ partial Latin Squares, of order $n$,
which satisfy the following properties:
\begin{enumerate}
\item $L \supset P_{m} \supset P_{m-1} \supset \dots \supset P_{2} \supset P_{1}=A$
\item $\forall i$, $1 \leq i \leq m-1$, $P_{i} \cup \{(r_{i},c_{i};e_{i})\}=P_{i+1}$
such that one of $P_{i} \cup \{(r_{i},c_{i};e_{i}\}$ or 
$P_{i} \cup \{(r_{i},c;e_{i}\}$ or $P_{i} \cup \{(r,c_{i};e_{i}\}$ is not a partial
Latin Square for any $e\in N / \{e_{i}\}$or $c\in N / \{c_{i}\}$ or $r\in N /
\{r_{i}\}$ respectively.
\end{enumerate}
\end{defn}

\subsection{The Proposed Scheme}
In Cooper \cite{cpr} a secret sharing scheme is constructed with a secret key
made from a Latin Square $L$, of order $n$. Furthermore \cite{cpr} notes the
following characteristics:
\begin{itemize}
\item The Latin Square $L$ is kept private, but its order however is made public.
\item The Shares are based on a partial Latin Square $S=\{ \cup A_{i}| A_{i} \in
L\}$ where $A_{i}$ is a critical set. With the union is taken over all possible critical sets in $L$ over some subset
of critical sets.
\item The number of critical sets used depends on the size of the Latin Square
and the number of shares.
\item The access structure is defined as $\Gamma = \{ B|B \subseteq \mathcal{S}
\; \& A \subseteq B\}$ where A is some critical set in $L$. Where $\Gamma$ is monotone
\end{itemize}
We shall now outline the basic protocol presented by Cooper \cite{cpr}:
\begin{itemize}
\item A Latin Square $L$ of order $n$ is chosen. The number $n$ is made public, but the Latin Square $L$ is kept secret and taken to be the key.
\item The set $S$ which is the union of a number of critical sets in $L$ 
\item For each $(i,j;k) \in S$, the share $(i,j;k)$ is distributed privately to a participant.
\item When a critical set of shares are brought together, they can reconstruct
the Latin Square $L$ and thus the secret key.
\end{itemize}

\subsection{The Ranking Problem}
The constructions proposed by \cite{cpr,aps,bean}, are such that each user is
given one element from a Latin Square and a subset of these elements may be
combined to form a critical set. In Donovan \cite{dcns}, a more general
construction is given such that, a set $S$ is the union of a number of critical
sets in a Latin Square. Elements from the set $S$ are dealt out to each player,
so that a group of players wish to reconstruct the critical set and the secret
can be recovered. This gives rise to the question to that complex issue in Latin
Squares of there being some positions which are more important than others.
\\
An intruder who knew C's share and the location
of the other shares, what the player did next would depend upon their knowledge
of the concurrence scheme.  If our player knew the scheme then one would 
start by guessing at two of the other shares (A and B, or D and E,
or A and D) in which case it is an disadvantage compared to an
intruder who knows a share other than C's.
\\
If our player does not know the scheme, it would seem most logical to try
to guess D's share before trying to guess two other shares at once.
Again, in this case, our player is at a disadvantage compared to an
intruder who knows a share other than C's.

\subsection{Security of a Latin Square Based Scheme}
The main security issues with this type of scheme were investigated 
heavily by Cooper \cite{cpr}. We shall now examine these vulnerabilities:
\begin{itemize}
\item An unauthorized players knows one $n$th of the critical set.
\item A group of unauthorized players have a greater chance of reconstructing
the critical set with their group of shares.
\item The security of this scheme is based on the number of possible latin
squares which contain the partial Latin Square defined by a disloyal group of
players. It has been estimated that the number of Latin Squares containing the
set $C$ for $\{(i,j;k)\}$ such that for a square of order $n=11$, $\geq 19000000$
\end{itemize}

\noindent The complexity of completing partial Latin Squares has been
investigated by Colbourn \cite{col}. The computational complexity of this problem
is NP-Complete. However even for a Latin Square of order $n=11$ there are still a
measurable number of solutions which can be generated by brute force. 

\section{RSA Threshold System}
Threshold schemes however are by no means perfect despite their proponents 
\cite{shoup}. Many of these schemes have a great
many short falls which include at least one of the following:
\begin{enumerate}
\item The scheme has no rigorous security proof
\item Share generation and verification is interactive and requires synchronous
communications network
\item The size of each share increases linearly with respect to the number of
players.
\end{enumerate}
In an effort to rectify this situation \cite{shoup} presents a new RSA threshold
scheme that exhibits the following:
\begin{enumerate}
\item Unforgeable and robust if we assume that the RSA problem is hard
\cite{riv}
\item Share generation and verification is completely non-interactive
\cite{riv}
\item The size of the share is bounded by a constant and the size of the
discrete logarithm problem \cite{sham} and \cite{men}
\end{enumerate} Shoup \cite{shoup} further stresses the fact that the share is a standard RSA
signature. This is underpinned by the fact that the public key and verification
algorithm are the same as for an RSA signature \cite{sham,riv}.
The refined model examined in this paper and in \cite{shoup} where there is one
threshold $t$ for the maximum number of traitors and $k$ is the minimum quorum
size.

\subsection{The RSA threshold Scheme}
We must first establish a set of players $w$, denoted $1,\ldots, w$, a trusted
designer/dealer, and traitor. This systems also has a signature verification,
a share verification and share combining algorithms. Shoup \cite{shoup} only
uses 2 variables, however in our investigation we must remain consistent with
the majority of the literature and consider 3 parameters. So we denote the number of corrupted players as $c$, the number of shares needed to produce a
signature as $t$ and the set of all users $w$. We also mention the requirement
for these parameters is, $t \geq c+1$ and $w-c \geq t$.
\\
The dealing phase is initiated by the dealer generating a public key, along
with a set of secret key shares and a set of verification keys. The corrupt
player obtains the secret key shares of the corrupted players, the public and
verification keys. The post dealing phase is when the corrupt player acts by
submitting a signing request to the loyal players for a message. After the
request has been submitted, a player outputs a signature share for the submitted
message.
\\
The signature verification algorithm takes an input message, a signature and a 
public key to determine if the signature is valid. The signature 
verification algorithm takes an input message, and  signature share on that message from player $i$, to
determine if that signature share is valid. The share combining algorithm takes
a message and $t$ valid signature shares on the message with the public key and
the verification keys. The algorithm then outputs a valid signature on that
message.
\\
The non-forgeability of signatures protocol dictates that if an adversary forges a
signature at the end of the protocol our player outputs a valid signature on a
message that was not submitted as a signing request to at least $t-c$ loyal
players. Furthermore we must stress that the threshold signature scheme is 
non-forgeable if it is computationally infeasible for the corrupt adversary to forge
a signature.

\subsection{Security of RSA Threshold}

\begin{thm}
For $t=w+1$, in the random oracle model for $H'$, the above protocol is a secure
threshold signature scheme which is robust and non-forgeable. Thus we assume that
the standard RSA signature scheme is secure.
\end{thm}

\noindent We shall only outline a very short comment on the proof for this theorem. One
should consult Shoup \cite{shoup} for a more detailed approach.
The robustness of the threshold signature scheme is cemented in its
non-forgeable. We assume that the standard RSA signature scheme is secure
because of the difficulty in solving the adaptive message attack. This statement
can be justified by the random oracle model of \cite{men} such that given some
random $x \in \mathbf{Z}^{*}_{n}$, it is hard to compute $y$ such that $y^{e}=x$

\section{Analysis}
We shall put forward the merits of a Latin Square SSS and the RSA 
based system to examine

\begin{itemize}
\item A Latin Square Scheme, can provide good security when the critical set on
which the scheme is founded is not based on strong critical or semi-strong
critical partial Latin Squares.
\item Latin Squares of large order i.e. $\geq 11$ provide for a relatively
secure system.
\item The current literature believes that the RSA problem is hard to compute
\item The Decision Diffie-Hellman (DDH) assumption-given some random $g, \; h
\in Q_{n}$, along with $g^{n}$ and $h^{b}$, it is hard to decide if $a \equiv b
\; mod \; m$.
\item Finding a correct authorized group of shares from one given share is
computationally difficult.
\end{itemize}

\noindent If we were to look at a computational attack against the Latin Square Scheme,
one would need only to find one disloyal player and simply generate a
completion for that share \cite{bean}. Although the prospect of finding a
solution to this problem becomes more difficult as the size of the scheme
increases beyond 11 players \cite{don}, it is still possible. Without a scheme
that allows for a disenrollment procedure \cite{don}, a brute force attack for
computing the completion of the Latin Square is a viable attack.
\\
If one already holds one of the other share then, there is a $1$ in $4$ 
chance of completing the critical set and discovering the secret by 
simply picking one share at random \cite{cpr}. A $25$ percent chance of
completing a critical set given one player is disloyal, is a risk not worth
taking in our view. If one player were somehow compelled or convinced that
becoming disloyal was appropriate then a scheme that placed so much trust in one
player is too risky. 
\\
Although the Latin Square model is entirely theoretical, it must be asked why
one would use such a scheme that has two major faults. Unless one can ensure
that no players will defect and become disloyal, then this scheme is far from
desirable. In contrast RSA based protocols are one of the best methods available
to ensure the security of a multiparty scheme for digital signatures
\cite{men,shoup}.

\end{document}